\documentclass[aps,prb,twocolumn,superscriptaddress]{revtex4-1}
\usepackage{graphicx}
\usepackage{amsmath}
\usepackage{epstopdf}

\usepackage[colorlinks=true,
            linkcolor=blue,
            urlcolor=blue,
            citecolor=blue]{hyperref}

\usepackage[abs]{overpic}
\usepackage{subfigure}
\usepackage[dvipsnames,svgnames]{xcolor}

\begin{document}

\title{Vibrational modes of negatively charged silicon-vacancy centers in diamond from \\{\it ab initio} calculations}

\author{Elisa Londero} 
\affiliation{Institute for Solid State Physics and
Optics, Wigner Research Center for Physics, Hungarian Academy of Sciences, P.O.\
Box 49, Budapest  H-1525, Hungary}

\author{Gerg\H{o} Thiering} 
\affiliation{Institute for Solid State Physics and
Optics, Wigner Research Center for Physics, Hungarian Academy of Sciences, P.O.\
Box 49, Budapest  H-1525, Hungary}
\affiliation{Department of Atomic Physics, Budapest
University of Technology and Economics, Budafoki \'ut 8, Budapest H-1111,
Hungary}

\author{Lukas Razinkovas}
\affiliation{Center for Physical Sciences and Technology (FTMC), Vilnius LT-10257, Lithuania}

\author{Adam Gali} \email{gali.adam@wigner.mta.hu} 
\affiliation{Institute for Solid State Physics and
Optics, Wigner Research Center for Physics, Hungarian Academy of Sciences, P.O.\
Box 49, Budapest  H-1525, Hungary}
\affiliation{Department of Atomic Physics, Budapest
University of Technology and Economics, Budafoki \'ut 8, Budapest H-1111,
Hungary}

\author{Audrius Alkauskas} \email{audrius.alkauskas@ftmc.lt}
\affiliation{Center for Physical Sciences and Technology (FTMC), Vilnius LT-10257, Lithuania}
\affiliation{Department of Physics, Kaunas University of Technology, Kaunas LT-51368, Lithuania}

\date{\today}

\begin{abstract} 
Silicon-vacancy (SiV) center in diamond is a photoluminescence (PL) center with a characteristic zero-phonon line energy at 1.681~eV that acts as a solid-state single photon source and, potentially, as a quantum bit. The majority of the luminescence intensity appears in the zero-phonon line; nevertheless, about 30\% of the intensity manifests in the phonon sideband. Since phonons play an essential role in the operation of this system, it is of importance to understand the vibrational properties of the SiV center in detail. To this end, we carry out density functional theory calculations of dilute SiV centers by embedding the defect in supercells of a size of a few thousand atoms. We find that there exist two well-pronounced quasi-local vibrational modes (resonances) with $A_{2u}$ and $E_u$ symmetries, corresponding to the vibration of the Si atom along and perpendicular to the defect symmetry axis, respectively. Isotopic shifts of these modes explain the isotopic shifts of prominent vibronic features in the experimental SiV PL spectrum. Moreover, calculations show that the vibrational frequency of the $A_{2u}$ mode increases by about 30\% in the excited state with respect to the ground state, while the frequency of the $E_u$ mode increases by about 5\%. These changes explain experimentally observed isotopic shifts of the zero-phonon line energy. We also emphasize possible dangers of extracting isotopic shifts of vibrational resonances from finite-size supercell calculations, and instead propose a method to do this correctly. 
\end{abstract}


\maketitle


\section{Introduction}
\begin{figure}
\includegraphics[width=8.5cm]{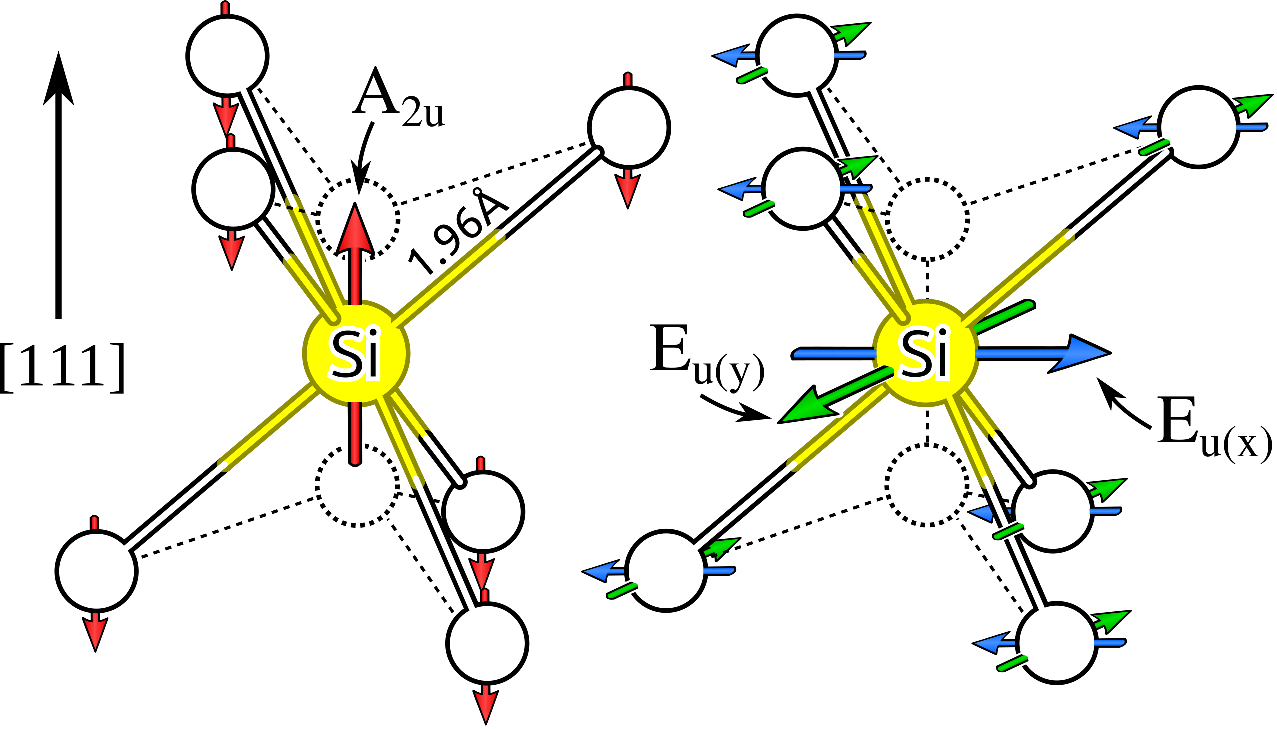}
\caption{(color online) 
The geometry sketch of the SiV center. The localized vibration modes $A_{2u}$ (motion of Si-atom parallel to $\langle 111 \rangle$ axis, left figure) and $E_{u}$ (motion of Si-atom perpendicular to $\langle 111 \rangle$ axis, right figure) are shown. These quasi-local vibration modes have significant localization on the six neighbor carbon atoms that move in the opposite direction to that of Si atom in the corresponding vibrations.
}
\label{SiV_fig}
\end{figure}

The study of optically active paramagnetic defects in diamond has a rich history \cite{Davies1981}. In the past two decades, nitrogen-vacancy (NV) centers attracted a lot of attention due to their potential applications in emerging {\it quantum technologies}, e.g., as spin qubits, sensors, or single-photon emitters \cite{Gordon2013}. More recently, the diamond silicon-vacancy (SiV) center has been demonstrated as a single photon source \cite{Neu2011,Sipahigil2014}. In pure diamond samples the zero-phonon-line (ZPL) of the SiV center is at 1.681~eV, thus, in the infra-red region. The defect is a $S=1/2$ center and it can be coherently manipulated optically at a single-site level \cite{Rogers2014}. Earlier studies indicated that the SiV center is negatively charged and that the Si atom resides in the symmetric split-vacancy configuration that exhibits $D_{3d}$ symmetry\cite{Goss1996} (Fig.~\ref{SiV_fig}), with defect oriented along the $\langle 111 \rangle$ axis. \emph{Ab initio} calculations found that the optical signal of the defect can be described by the transition between $e_u$ and $e_g$ defect states \cite{Gali2013}. Both the ground $^2E_g$ (electronic configuration $e_u^4e_g^3$) and the excited $^2E_u$ ($e_u^3e_g^4$) states are dynamic Jahn-Teller (JT) systems~\cite{Hepp2014,Rogers2014c} that preserve the high $D_{3d}$ symmetry \cite{IRREP}. 

Like for other defects, lattice vibrations play an important role in defining many properties of SiV centers. Most prominently, this concerns optical properties: while luminescence of SiV centers is mainly dominated by the ZPL transition with $\sim$70\% of the total emission, the remaining $\sim$30\% show up in the phonon sideband with a few distinct phonon replicas \cite{Sternschulte1994}. In a recent study one of those replicas, occurring at 64~meV from the ZPL energy and having a relative intensity of $\sim$3\% w.r.t.\ the ZPL, was shown to exhibit an isotope shift that was proportional to the inverse square-root of the silicon mass for three different Si isotopes ($^{28}$Si, $^{29}$Si, $^{30}$Si) \cite{Dietrich_NJP_2014}. In absolute units, downward shifts for $^{29}$Si and $^{30}$Si isotopes were 8.22 cm$^{-1}$ (1.02 meV) and 17.81 cm$^{-1}$ (2.21 meV). Earlier density functional theory (DFT) calculations \cite{Goss_PRB_2007} indeed found a quasi-local mode with similar energy (56.5 meV), but with significantly smaller calculated shifts of 5 and 10 cm$^{-1}$, respectively, calling for a deeper analysis of quasi-local SiV vibrations. Phonons also play an important role in nonradiative processes in defects. It is known that quantum efficiency of optical emission at SiV centers most likely does not exceed 20 \%, indicating the existence of detrimental non-radiative decay channels \cite{Rogers2014b}. Therefore, phonons affect or govern both radiative and non-radiative processes at SiV centers, but a deeper understanding of them is currently missing. In this paper, we report \emph{ab initio} density functional theory (DFT) calculations of vibrational modes of effectively isolated SiV in both ground and excited states. Calculations are accompanied by the group theory analysis of the phonon spectrum of the SiV center. Our findings explain a few recent experimental observations. Also, they solve the contradiction between calculated isotope shifts reported in Ref.~\onlinecite{Goss_PRB_2007} and those measured in Ref.~\onlinecite{Dietrich_NJP_2014}.

The paper is organized as follows. In Sec.~\ref{methods} we present the details of DFT calculations, including the electronic structure methods, calculations of vibrational modes, and analysis of vibrations. In Sec.~\ref{results} we present the results of calculations of vibrations in the ground state, as well as discuss isotopic shifts of phonon modes. In Sec.~\ref{results_e} the results for the excited electronic state are presented. In Sec.~\ref{disc} we discuss our calculations in light of experiments and highlight remaining unanswered questions. Finally, in Sec.~\ref{summ} we summarize our work and draw conclusions. Throughout the paper negatively charged SiV$^-$ is labeled simply SiV.


\section{Methodology \label{methods}}

\subsection{Electronic structure and vibrational spectra}

{\it Ab initio} calculations employed in the current work were similar to the previous study \cite{Gali2013}. In short, we determined the electronic structure of the SiV center within the framework of DFT as implemented in the \textsc{vasp} code \cite{VASP1}. We used the hybrid Heyd-Scuseria-Ernzerhof functional \cite{HSE} HSE06 (HSE for short) with a standard fraction of screened Fock exchange $a=0.25$. Lattice constant $a=3.544$~\AA\ and the band gap $E_g=5.3$~eV are in excellent agreement with experimental values. The use of a hybrid functional for the SiV center is especially important to describe the excited state \cite{Gali2013}.  We used the projector-augmented wave approach with a plane-wave energy cutoff of 400 eV \cite{PAW,VASP2}. The Brillouin zone was sampled at the $\Gamma$ point, and this choice was made in order to make sure that the local symmetry is correctly described. We used a 216-atom cubic supercell for actual defect calculations \cite{Freysoldt2014}. The $^2E_u$ excited state was calculated by the constrained-DFT method (originally due to Slater \cite{Slater}, see Ref.~\onlinecite{Kaduk2012} for a review of new developments), whereby the electron is promoted from the $e_u$ level to the $e_g$ defect level. The methodology was previously applied to the NV center in diamond \cite{Gali2009}, where a very good agreement with experiment was demonstrated. 

The HSE functional provides an accurate description of both the geometry and the electronic structure of the SiV center \cite{Gali2013}. The calculated ZPL energy for the transition $^2 E_g\rightarrow{}^2E_u$ of 1.72~eV compares very favorably with the experimental value of 1.681~eV \cite{Gali2013}. Thus, it is desirable to also calculate vibrations at the hybrid functional level. However, this is prohibitively costly even for the 216-atom supercell. To calculate the vibrational of HSE-quality in very large supercells (essentially dilute limit) we use the embedding procedure applied and tested on the NV center in diamond in Ref.~\onlinecite{Alkauskas_NJP_2014}.

The procedure relies on the fact that the dynamical matrix in diamond is rather short-ranged \cite{SUPPL}. Dynamical matrix for a defect system was constructed as follows \cite{Alkauskas_NJP_2014}. If two atoms are further away than 4.2~\AA\ then the matrix element is set to zero. Otherwise, if {\it at least one} of the atoms is closer to the Si atom than 2.75~\AA\ then the dynamical matrix element is taken from the 216-atom HSE calculation. For all other atom pairs we used the value derived from bulk diamond calculations also performed in the 216-atom supercell. In this way we constructed dynamical matrices for supercells containing up to 5832 atoms. We label supercells $N\times N \times N$, where $N$ is the number of cubic diamond unit cells in any direction. Pristine supercells contain $8N^3$ atoms and thus supercells with $N= $ 3, 4, 5, and 6, 7, 8, and 9 contain the nominal number of atoms $M= $ 216, 512, 1000, 1726, 2744, 4096, and 5832, respectively. For bulk diamond, the resulting phonon spectrum is very close to the experimental one. E.g., the energy of the longitudinal optical mode at the $\Gamma$ point is 167~meV, in excellent agreement with the experimental value of 166.7 meV \cite{Warren1967}. 

In all defect calculations of the dynamical matrix we used $1/2$ occupations for spin-unpaired $e$ orbitals. Actual electronic configurations were $e_{ux}^{2}e_{uy}^{2}e_{gx}^{3/2}e_{gy}^{3/2}$ for the ground and $e_{ux}^{3/2}e_{uy}^{3/2}e_{gx}^{2}e_{gy}^{2}$ for the excited state. The electron density of such states corresponds to an average density of $E_x$ and $E_y$ states pertaining to $^2E_g$ or $^2E_u$ manifolds, respectively. This results in a $D_{3d}$ symmetry, allowing us to determine vibrational frequencies without complications associated with the Jahn-Teller effect. This approach is justified by experimental results that indicate $D_{3d}$ symmetry \cite{Hepp2014}. Vibronic coupling is not addressed in this work.

\subsection{Characterization of phonons}

All the vibrational modes have been characterized according to the irreducible representation of the D$_{\text{3d}}$ point group: $A_{1g}$, $A_{1u}$, $A_{2g}$, $A_{2u}$, $E_g$, and $E_u$ \cite{Landau}. This has been done by calculating scalar products of the type $(\Delta\mathbf{r}_k \cdot \hat{O}\Delta\mathbf{r}_k)$, where $\Delta\mathbf{r}$ is a short-hand notation for the eigenmode $k$, and $\hat{O}$ is a symmetry operation. The eigenmode $k$ is described by a vector with components $\Delta r_{k; \alpha i}$, where $\alpha$ labels atoms, and $i=\{x,y,z\}$. For a uniquely determination of the irreducible representation, it is sufficient to consider symmetry operations $C_3$ (rotation), $\sigma_d$ (reflection in the plane that contains the symmetry axis) and $i$ (inversion). 

Vibrations were also characterized by their localization. A bulk phonon is entirely delocalized and many atoms participate in that particular vibration. Point defects usually change the vibrational structure of a solid. Sometimes, the defect gives rise to a localized vibrational mode with a frequency outside the vibrational spectrum of the host material. However, defects can also induce quasi-local modes. Their frequencies (energies) overlap with the bulk phonon spectrum, and thus they are not strictly localized, but often they give rise to observable spectroscopic signatures. To quantify the localization, vibrational modes corresponding to each irreducible representation (we say below simply ``symmetry'') are characterized by the inverse participation ratio (IPR):
\begin{equation}
\text{IPR}_k = \frac{1}{\sum_\alpha p_{k;\alpha}^2} \text{,}
\label{IPR}
\end{equation}
where
\begin{equation}
p_{k;\alpha} = \sum_{i}\Delta r_{k; \alpha i}^2.
\end{equation}
The sum in the first equation runs over all atoms $\alpha$, and, as before, $i=\{x,y,z\}$. IPR essentially describes onto how many atoms the given mode is localized \cite{Bell_JPCC_1970,Alkauskas_NJP_2014}. E.g., when only one atom vibrates in a given mode, $\text{IPR}=1$. When all $M$ of the supercell vibrate with equal amplitudes, $\text{IPR}=M$. If every-second atom of the supercell vibrates with an equal amplitude, $\text{IPR}=M/2$; etc.

We also define a localization ratio $\beta_k$ \cite{Alkauskas_NJP_2014}:
\begin{equation}
\beta_k=M/\text{IPR}_k=M\sum_\alpha p_{k;\alpha}^2 \text{.}
\label{beta}
\end{equation}
$\beta_k$ quantifies the fraction of atoms (more precisely, the inverse of this fraction) in the supercell that vibrate for a given vibrational mode $k$. $\beta_k=1$ for modes when all atoms in the supercell vibrate with the same amplitude, while $\beta_k\gg1$ for quasi-localized and local modes.


\section{Results: ground state \label{results}}

We first focus on the vibrations in the electronic ground state $^2E_g$. In the current Section we study the $^{28}$Si isotope, and leave the study of isotopic shifts to Sec.~\ref{isotope}. At the outset, we note that calculations show no truly local vibrational modes with frequencies outside the bulk phonon band. This is expected, as chemical interactions are typically weaker in the vacancy environment and, furthermore, the Si atom is heavier than surrounding carbon atoms. Thus, our subsequent analysis is focused on bulk and quasi-local modes.

\subsection{IPR analysis of the vibration modes}

\begin{figure}
\includegraphics[width=8.0cm]{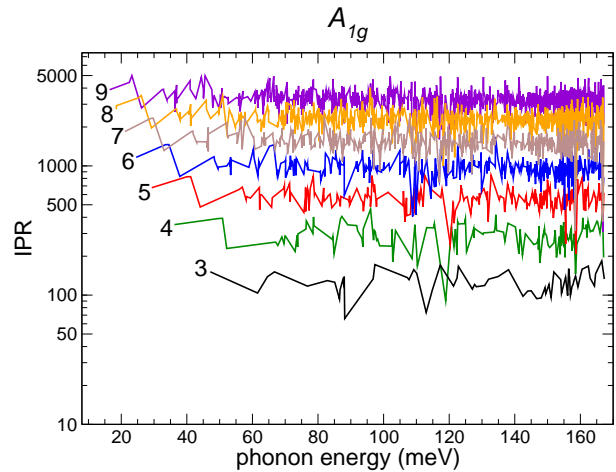}
\caption{(Color online) Inverse participation ratios [Eq.~(\ref{IPR})] of $A_{1g}$ symmetry modes for different supercell sizes as a function of energy. Systems with $N=3, 4, 5, 6, 7, 8, 9$ (size of a cubic supercell, a number next to each graph) correspond to bulk supercells containing $M$$=$$8N^3$ atoms. There are no quasi-local modes of $A_{1g}$ symmetry.
}
\label{a1g}
\end{figure}

Let us first start with $g$ (even, or {\it gerade}) modes that are symmetric with respect to inversion. Since the Si atom is at the center of inversion, Si atom does not participate in these even-parity vibrations. Fig.~\ref{a1g} shows calculated IPRs for fully symmetric $A_{1g}$ modes as a function of energy for different supercell sizes. $A_{1g}$ vibrations represent ``breathing'' motion with respect to the Si atom. We see that for a given supercell the IPRs of all vibrational modes are very comparable. In fact, they are all similar to IPRs of bulk modes calculated for the same supercell size. Small oscillations are due to variations of IPR values for phonons at different parts of the Brillouin zone. Energies of modes start with the smallest available mode in a given supercell and end with $\sim$167~meV. 

\begin{figure*}
\includegraphics[width=16.0cm]{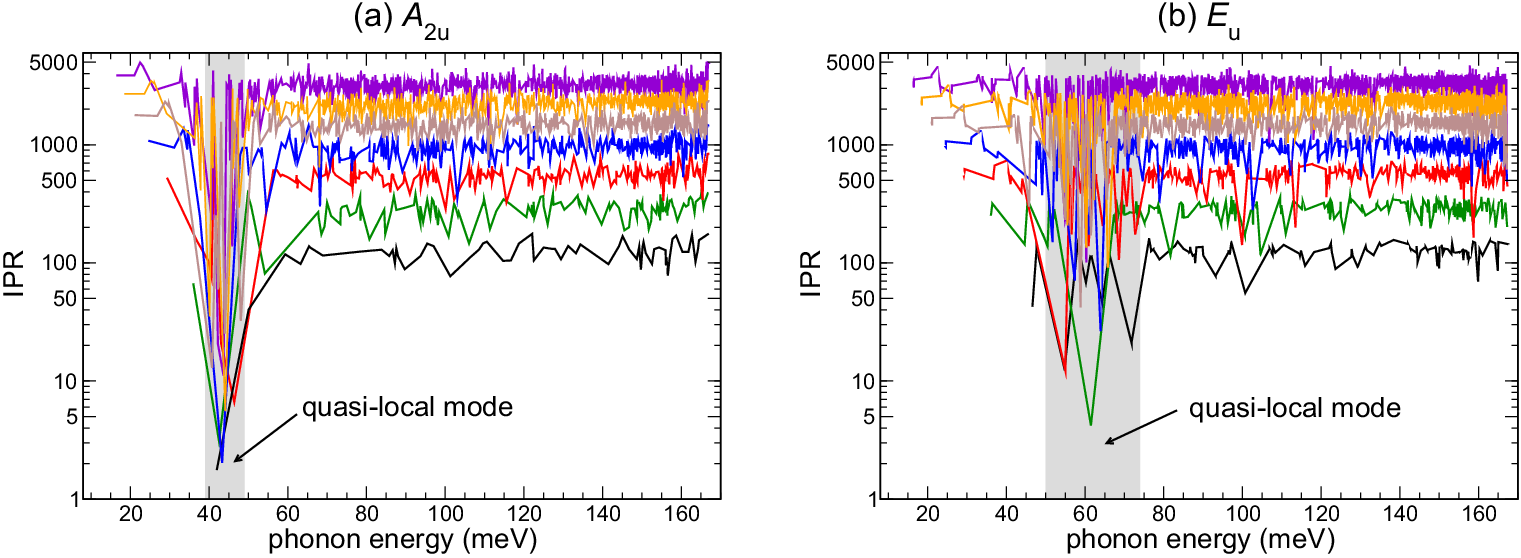}
\caption{(Color online) (a) Inverse participation ratios [Eq.~(\ref{IPR})] for $A_{2u}$ symmetry modes for different supercell sizes as a function of energy. (b) Same for $e_{u}$ modes. Supercells the same as in Fig.~\ref{a1g}.
}
\label{a2u}
\end{figure*}

Fig.~\ref{a1g} shows that there are no clearly pronounced quasi-local modes of $A_{1g}$ symmetry.  Those should have IPRs that are noticeably smaller than IPRs of other modes. IPRs exhibit some variation, but in general we conclude that $A_{1g}$-symmetry vibrations are just slightly perturbed bulk modes of diamond. The same conclusion holds for modes of $A_{2g}$ symmetry \cite{SUPPL}. In $D_{3d}$ symmetry $A_{2g}$ modes represent rotation motion around the defect axis \cite{Landau}. Interestingly, the IPR analysis of $E_g$ modes also shows no indication of the presence of quasi-local $E_g$ modes \cite{SUPPL}. This is an important conclusion, since the dynamic Jahn-Teller effect seen in SiV centers \cite{Hepp2014,Rogers2014c} preserves the inversion symmetry in the ground (as well as the excited state), and, consequently, couples to $E_g$ modes. These modes represent rotation motion around the two axes that are perpendicular to the defect axis. The absence of quasi-local modes of $E_g$ symmetry also shows the complex nature of the Jahn-Teller effect. In order to properly understand it, one needs to include the coupling to a continuum set of $E_g$ modes \cite{OBrien1972} that are essentially bulk-like. The investigation of the Jahn-Teller effect is beyond the scope of our study. 

Analysis above shows that there are no quasi-local modes of $g$ symmetry and we find only slightly perturbed bulk modes. Now we turn to the $u$ (uneven, or {\it ungerade}) modes that break the inversion symmetry. Si atom can contribute to the vibration of these modes. However, IPRs of $A_{1u}$ modes exhibit behavior similar to $A_{1g}$ and other $g$ vibrations: there are no quasi-local modes \cite{SUPPL}.

The situation is radically different for the $A_{2u}$ modes. In the $D_{3d}$ point group the $A_{2u}$ irreducible representation is special as the $z$ coordinate transforms according to it ($z$ being the symmetry axis). The IPR analysis is shown in Fig.~\ref{a2u}(a). We see a clearly pronounced quasi-local mode with energy $\sim$$(39-49)$~meV where IPRs are significantly smaller than the average IPR for a particular supercell (cf.~the situation with $A_{1g}$ modes shown in Fig.~\ref{a1g}). A dip happens irrespective of the size of the supercell. The actual frequencies of vibrational modes that have significantly smaller IPRs differ in different supercells, but this is expected due to finite-size effects. We also see in Fig.~\ref{a2u}(a) that there are no quasi-local modes of $A_{2u}$ symmetry at other energies. Analysis of the vibrational pattern for the quasi-local mode shows that in this vibrations it is predominantly the Si-atom that moves along the symmetry axis, as indicated in Fig.~\ref{SiV_fig}. 

Calculated IPRs for $E_u$ modes are shown in Fig.~\ref{a2u}(b). The $e_{u}$ irreducible representation is also special as the $\{x,y\}$ coordinates transform according to it. The IPR analysis indicates that there is a quasi-local mode with energies at $\sim$$(52-72)$~meV. Compared to the $A_{2u}$ mode, the $E_u$ resonance is broader, as the decrease of IPRs for the quasi-local mode with respect to delocalized modes is not as significant. The quasi-local vibration mostly corresponds to the Si atom vibrating perpendicular to the defect axis, as shown in Fig.~\ref{SiV_fig}.

The results of this Section show that the SiV induces quasi-local vibrations of $A_{2u}$ and $E_{u}$ symmetries. In the next Section we analyze these vibrations in more detail. 

\subsection{Localization ratios of $A_{2u}$ and $E_u$ modes \label{loc}}

\begin{figure*}
\includegraphics[width=16.0cm]{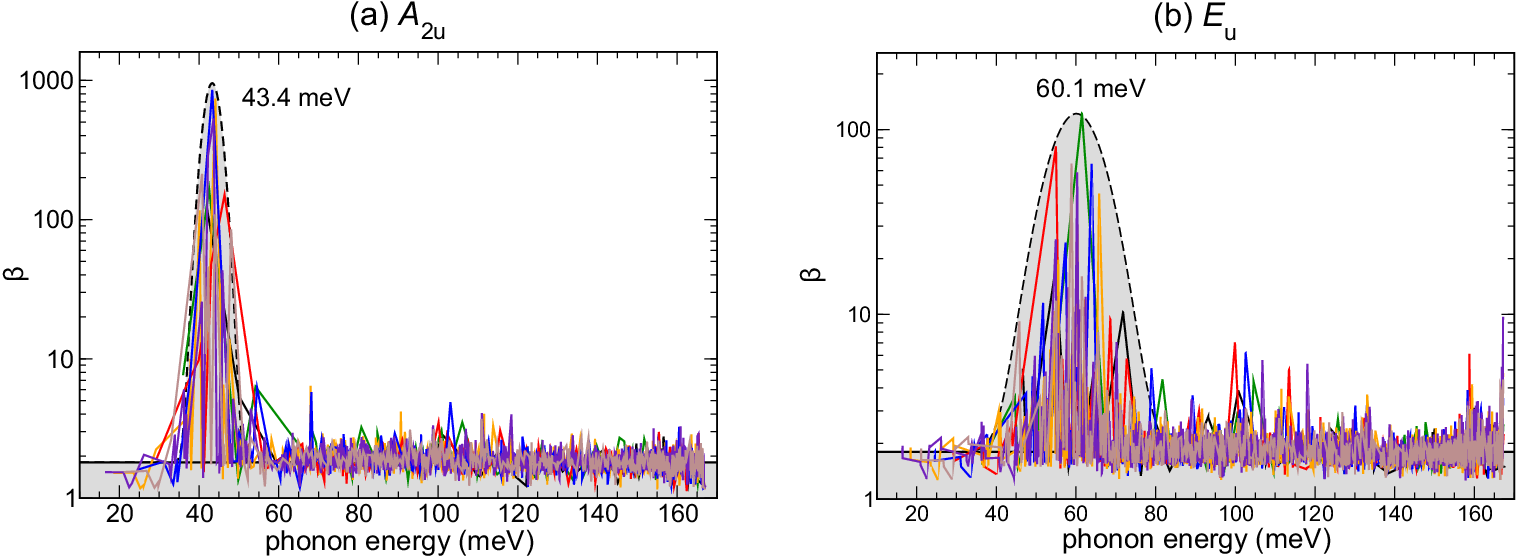}
\caption{(Color online) (a) Localization ratios [Eq.~(\ref{beta})] for $A_{2u}$ symmetry modes for different supercell sizes as a function of energy. Localization ratios are calculated from the data in Fig.~\ref{a2u}, and results from different supercells are superimposed on top of one another. The dashed line and shaded region serve as guides to the eye. (b) Same for $E_u$ modes. Note that the $y$ axis has logarithmic scale.}
\label{a2u-beta}
\end{figure*}

For an alternative analysis of quasi-localized modes we use the localization ratio $\beta$ [Eq.~(\ref{beta})]. Results for $A_{2u}$ modes are shown in Fig.~\ref{a2u-beta}(a). Different graphs represent calculations from different supercells, and here they fall on top of each other. For most of the vibrations $\beta$ fluctuates around a constant value, but one clearly sees the emergence of a quasi-local mode with $\beta \gg 1$. Analysis in terms of $\beta$ is only qualitative ($\beta$ is not a physical quantity that can be measured). However, in Fig.~\ref{a2u-beta}(a) we tentatively draw an envelope function to make the reading and understanding of our calculations results clear. This allows us to determine the energy of the $A_{2u}$ quasi-local vibration, about $\sim$$43$~meV. We will provide a more quantitative definition of this energy below. 

Analysis of the localization ratio of the $E_u$ modes is shown in Fig.~\ref{a2u-beta}(b). The emergence of the $E_u$ resonance is clearly seen. The energy of the $E_u$ vibration is $\sim$$60$~meV, and it can be seen that the $E_u$ resonance is broader than the $A_{2u}$ resonance.

\subsection{Isotope shifts of the quasi-local $A_{2u}$ and $E_u$ modes \label{isotope}}

Ref.~\onlinecite{Dietrich_NJP_2014} reported isotope shifts of the sharp feature in the phonon sideband $\Omega_{28}/\Omega_{29}=1.016$ and $\Omega_{28}/\Omega_{30}=1.036$ (Table \ref{Table2}), that closely correspond to the ``ideal'' shifts of $\sqrt{29/28}$=1.018 and $\sqrt{30/28}=1.036$ (i.e., $\Omega\sim$$1/\sqrt{m_\text{Si}}$). At low temperatures the phonon sideband in the luminescence spectrum reveals the vibronic structure of the electronic ground state, and thus it is justified to compare experimental results with calculations of the vibrational modes in the ground state. However, in this case we must be very cautious in comparing calculated frequencies of quasi-local vibrational modes with energies of features in the phonon sideband. This is because of a possible complex nature of this phonon sideband. We leave a discussion about comparison of the phonon sideband in luminescence for Sec.~\ref{disc}, and here only analyze the isotopic shifts of vibrations themselves.

\begin{figure}
\includegraphics[width=8.0cm]{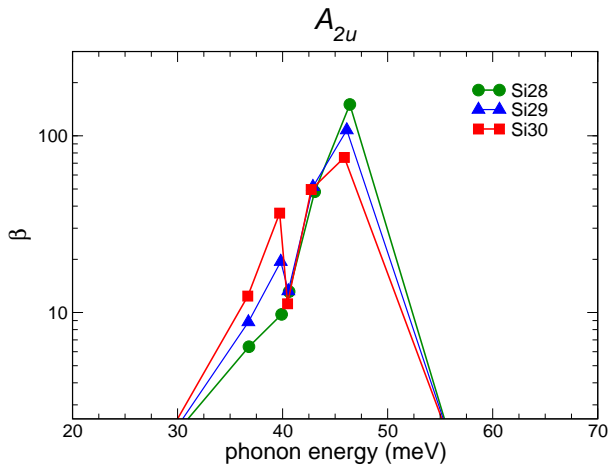}
\caption{(Color online) Localization ratios [Eq.~(\ref{beta})] for the $A_{2u}$ (a) symmetry quasi-local modes around energy 43~meV for three Si isotopes. Calculations performed for the $5\times 5\times 5$ supercell.
}
\label{a2u-beta-shift}
\end{figure}

In Fig.~\ref{a2u-beta-shift} we present the calculated localization ratios of the $A_{2u}$ vibrational resonance in the energy range $20-70$ meV for the $5\times 5\times 5$ supercell (nominally containing 1000 atoms) for Si isotopes with $m_{\text{Si}}=28$, 29, and 30~a.u. The calculation with the Si mass of 28~a.u. is the same, as presented in Fig.~\ref{a2u-beta}(a) for the corresponding supercell size. The quasi-local mode has a certain width, meaning that it is made of an infinite number of vibrations with energies around a peak energy of the quasi-local mode. As we deal with finite supercells, this means that for a specific supercell a quasi-local mode splits into a finite number of vibrations. We consider that the vibration for a chosen energy window contributes to the quasi-local mode if its localization ratio is significantly larger than the average localization ratio of all vibrations in the supercell. The criteria ``chosen energy window'' and ``significantly larger'' are motivated by the results shown in Fig.~\ref{a2u-beta}(a), and we choose the window $20-70$ meV and $\beta>3$.

For the particular supercell $5\times 5\times 5$ we can identify five specific vibrations that contribute to the quasi-local mode for all three Si isotopes. The vibration with the largest localization has energies 46.40, 46.11, and 45.88~meV for $^{28}$Si, $^{29}$Si, and $^{30}$Si, respectively. If we were to take these values without any further consideration, we would get isotope shifts $\omega_{28}/\omega_{29}=1.006$ and $\omega_{28}/\omega_{30}=1.011$, much smaller than the experiment values of 1.018 and 1.036 \cite{Dietrich_NJP_2014}. However, taking calculated frequencies face value would be valid for a truly localized mode outside the bulk phonon spectrum but is certainly incomplete for a quasi-local mode. 

As can be clearly seen in Fig.~\ref{a2u-beta-shift}, as we increase the mass of the isotope, two things happen. First, there are shifts of energies pertaining to the individual modes, as just discussed. Localization (quantified in terms of $\beta$) of higher-frequency modes tend to decrease, while localization of lower-frequency modes tend to increase. If we associate $\beta$ with the weight of the contribution of each specific vibration to the quasi-local mode, we see that there is a re-distribution of weight from the modes with larger energies to the modes with smaller energies. To take this effect into account, we can determine the energy of the quasi-local mode $\Omega$ as the weighted average of all the modes that contribute to this resonance:
\begin{equation}
\Omega=\frac{\sum_k\beta_k\omega_k}{\sum_k\beta_k}.
\label{ql}
\end{equation}
While there is freedom in choosing the weight, localization ratio $\beta_k$ is our first choice. If we now weigh the contribution of various modes, we get average energies 44.81, 43.90, and 42.88~meV for the Si isotopes 28, 29, and 30~a.m.u. This now corresponds to isotope shifts $\Omega_{28}/\Omega_{29}=1.021$ and  $\Omega_{28}/\Omega_{30}=1.045$. One finding is clear: actual isotope shifts for quasi-local vibrations are different than the estimates based on the analysis of individual vibrational modes. 

\begin{figure}
\includegraphics[width=8.5cm]{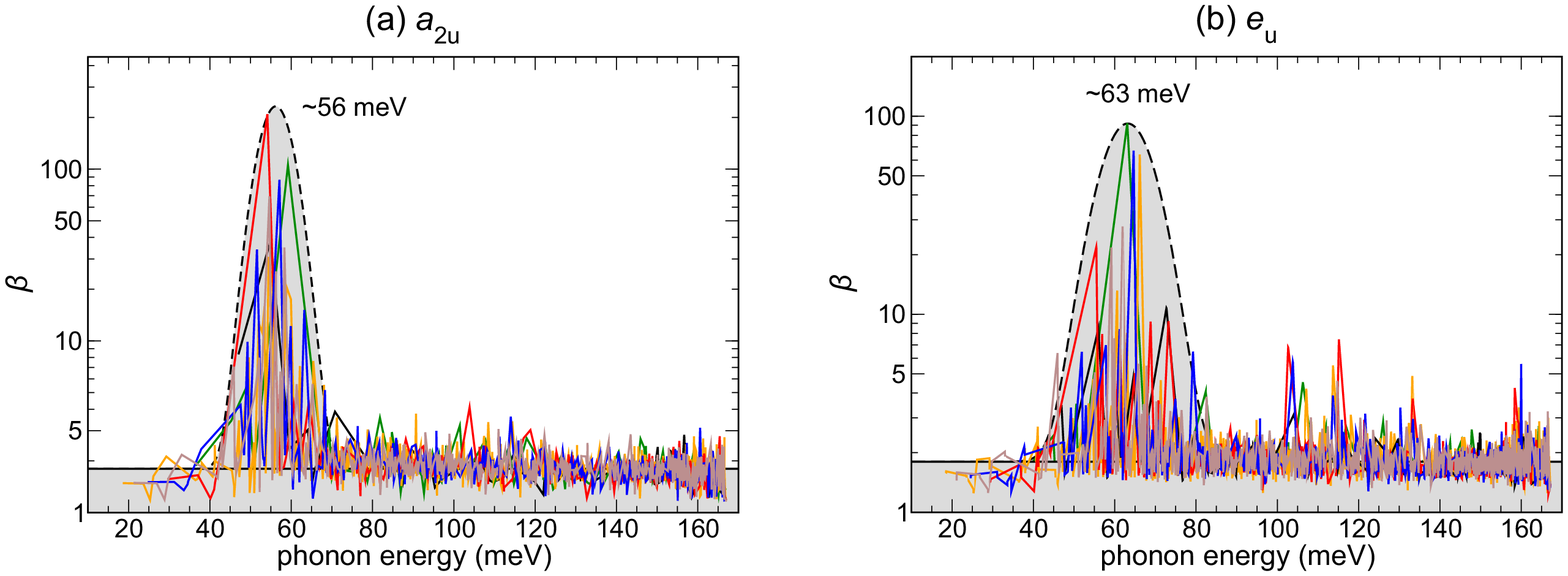}
\caption{(Color online) Frequencies of $A_{2u}$ and $E_u$ vibrational resonances [Eq.~(\ref{ql})] as a function of supercell size $N$. ``Bare value'' (black circles) is the frequency for a given supercell. ``$M$-average'' (red squares) and ``$N$-average'' (blue triangles) are weighted averages, as described in the main text. Horizontal dashed line corresponds to the extracted energy of the resonances.
}
\label{conv}
\end{figure}

We performed a similar analysis for all the considered supercells. The results for the $A_{2u}$ mode are shown in 
Fig.~\ref{conv}, top panel (``bare value'', black circles), which shows the dependence of the frequency of the $A_{2u}$ mode on the supercell size. We find that there is still a cell-to-cell variation of the resulting frequency even for the largest supercells, where even--odd oscillations of decreasing amplitude can be identified. These oscillations stem from the change in the effective sampling of the Brillouin zone as the size of the supercell increases. Since this effective sampling is denser for larger supercells, it is compelling to associate the frequency of the vibrational resonance with a weighted average, whereby the weight is proportional to the size of the supercell $M$. The dependence of the average frequency calculated in this way as a function of supercells included in the averaging is shown in Fig.~\ref{conv}, top pannel (``$M$-average'', red squares). As expected, the results converge to the final result much quicker. Convergence is even faster if the contributions of the supercells are weighted by $N$ (``$N$-average'', blue triangles). It is assuring, however, that the two averages converge to nearly the same value. In this way we obtain converged frequencies of the $A_{2u}$ resonance 43.38, 42.62, and 41.91~meV for the three Si isotopes. Given the nature of convergence (even--odd oscillations), we extract the error bar of $0.10$~meV pertaining to the procedure of averaging. This corresponds to the maximum difference between average frequencies from the last three supercells. This error is smaller than the inherent error of DFT calculations.
Calculated frequencies for the three isotopes correspond to isotopic shifts of $\Omega_{28}/\Omega_{29}=1.018$ and $\Omega_{28}/\Omega_{30}=1.035$, almost exactly the ``ideal'' isotopic shifts and thus very close to experimental values. Since we associate $\beta$ with the weights of different specific vibrations, we can also define the width $w$ of the quasi-local mode as:
\begin{equation}
w^2 = \frac{\sum_k \beta_k \left (\omega_k-\Omega \right )^2}{\sum_k\beta_k}.
\label{wql}
\end{equation}
$A_{2u}$ resonance is very narrow, and its calculated average width $w$ is only $1.9$~meV. The width defined above is a qualitative estimate useful to compare different vibrations, but the actual value should of course be taken with some caution. 

By reiterating the procedure for the quasi-local $E_u$ mode (Fig.~\ref{conv}, bottom panel) we obtain energies of vibrational resonances 60.00, 58.82, and 57.76 meV for the three Si isotopes. This gives $\Omega_{28}/\Omega_{29}=1.021$ and  $\Omega_{28}/\Omega_{30}=1.039$. Like for the $A_{2u}$ quasi-local mode, isotopic shifts are close to the ``ideal'' shifts. The calculated width of the $E_{u}$ quasi-local mode is $w=6.1$~meV, thus, larger than for the $A_{2u}$ mode. Nearly ``ideal'' isotopic shifts for both $A_{2u}$ and $E_u$ modes allow to think of them as representing the motion of an isolated Si-atom inside a hard diamond cage. $A_{2u}$  mode corresponds to vibration along the $z$ axis, while $E_{u}$ modes correspond to vibrations in the $xy$ plane (cf.~Fig.~\ref{SiV_fig}). Energies and widths of these quasi-local modes are summarized in Table \ref{Table2}. 

The analysis of this Section explains why earlier first-principles calculations of isotope shifts related to quasi-local modes \cite{Goss_PRB_2007} were smaller than the experimental values \cite{Dietrich_NJP_2014}. We showed that in order to calculate those shifts one needs to include all the vibrations that contribute to a specific resonance, e.g., via the use of Eq.~(\ref{ql}). The results for both $A_{2u}$ and $E_u$ resonances are summarized in Table \ref{Table2}.

\begin{table}[b]
\caption{Main results of this work compared with experiment. Theory: calculated frequencies $\Omega$, widths $w$ (both for the $^{28}$Si isotope), and isotopic shifts of quasi-local $A_{2u}$ and $E_{u}$ modes in the ground and excited states. Values of frequencies (widths) are given to two (one) significant digits. Experiment: energy $\Omega$, width and isotopic shifts of vibrational sidebands in the photoluminescence (PL) spectrum. All experimental data from Ref.~\onlinecite{Dietrich_NJP_2014} except where indicated. ND = ``not detectable''.}

\begin{ruledtabular}

\begin{tabular}{c c c c c c}
\multicolumn{6}{c}{Theory: vibrational resonances} \\
\hline
State	            &symmetry	&$\Omega_{28}$ (meV)	& $w$ (meV) & $\Omega_{28}/\Omega_{29}$ & $\Omega_{28}/\Omega_{30}$ \\
\hline
$^2E_g$	&$A_{2u}$   &43.4	& 1.9  & 1.018  & 1.035  \\
ground  &$E_u$	    &60.1	& 6.1  & 1.021  & 1.039  \\
\hline
$^2E_u$	 &$A_{2u}$  &56.1	& 3.9  & 1.016  & 1.033  \\
excited  &$E_u$	    &63.2	& 6.5  & 1.017  & 1.036  \\
\hline
\hline
\multicolumn{6}{c}{Experiment: PL phonon sidebands} \\
\hline
State	  &symmetry	         & $\Omega_{28}$ (meV) & $w$ (meV)& $\Omega_{28}/\Omega_{29}$ & $\Omega_{28}/\Omega_{30}$ \\
\hline
$^2E_g$ &$A_{2u}$ (Ref.~\onlinecite{Rogers2014c})& 63.8 &$\sim$$5$   & $1.016$    & $1.036$                   \\	
        &                                        & 42   & $>$$25$    & ND         & ND                        \\

\end{tabular}
\label{Table2}

\end{ruledtabular}
\end{table}

\section{Results: excited state \label{results_e}}

\begin{figure*}
\includegraphics[width=16cm]{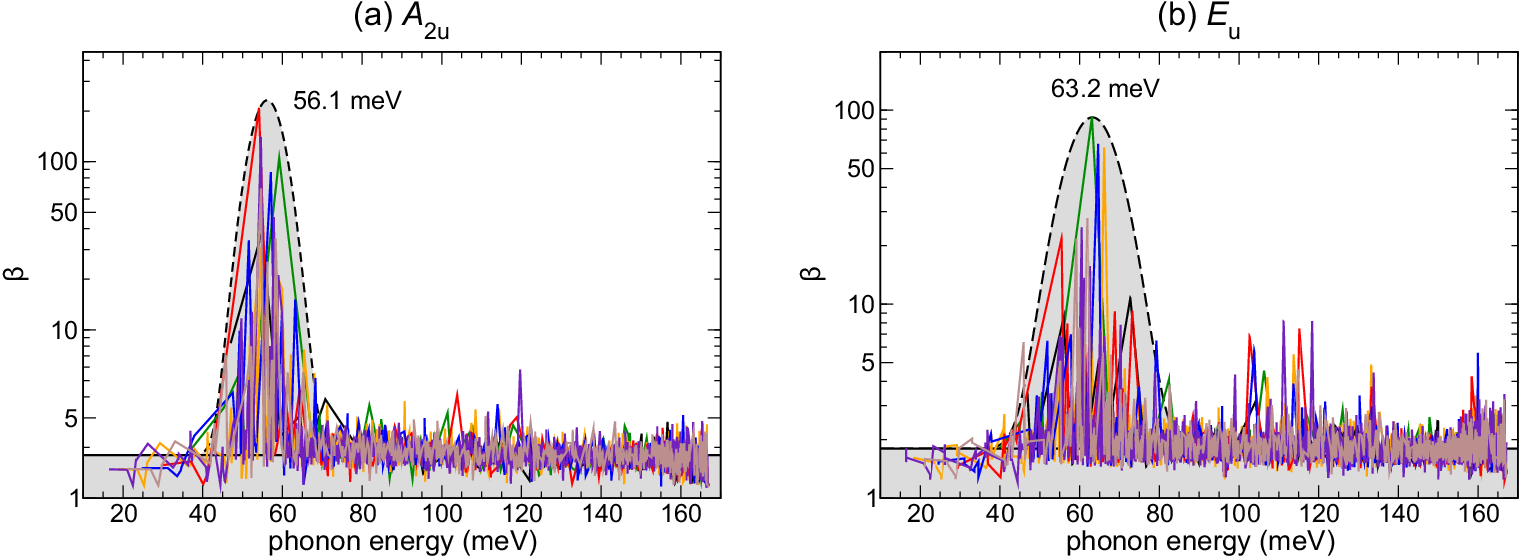}
\caption{(a) Localization ratios [Eq.~(\ref{beta})] for $A_{2u}$ (a) and $E_{2u} (b)$ symmetry modes on the excited state for different supercell sizes as a function of energy. The dashed line and shaded region serve as guides to the eye. The $y$ axis has logarithmic scale.}
\label{a2u-beta-e}
\end{figure*}

We have performed similar analysis for the vibrational modes in the electronic excited state $^2E_u$. Regarding all $g$ modes and $A_{1u}$ vibrations, the results are similar to the ground state: there are no quasi-local vibrations. However, analogously to the ground state, we find a very pronounced $A_{2u}$ resonance. Calculated localization ratios $\beta$ are shown in Fig.~\ref{a2u-beta-e}(a). Importantly, the energy of the resonance is 56.1 meV, about 30\% larger than in the ground state. Compared to the ground state, the $A_{2u}$ resonance is broader in the excited state with $w=3.9$~meV (Table \ref{Table2}), reflecting the increase of the bulk density of vibrational states for this energy \cite{Warren1967}. We also find a pronounced $E_{u}$ resonance, as shown in Fig.~\ref{a2u-beta-e}(b). The energy of the resonance is 63.2 meV, thus, increased by about 5\% with respect to the ground state, and its width, $w=6.5$ meV, is only slightly larger than for that of the ground state $E_u$ resonance. We find that excited state resonances also exhibit nearly ``ideal'' isotopic shifts (Table \ref{Table2}).

The increase of vibrational frequencies in the excited state vs.~the ground state can be understood from the electronic structure of the SiV center. The optical excitation corresponds to removing the electron from a more delocalized $e_u$ orbital to a more localized $e_g$ orbital \cite{Gali2013}. This increases the electron density around the Si atom, and also thus the vibrational frequencies. 

We end the analysis of the vibrations by taking a critical look at our computational setup, in which the Jahn-Teller effect was not taken into account because of the use of partially-occupied $e_g$ electronic states (see Sec.~\ref{methods}). As discussed earlier, the Jahn-Teller effect seen in SiV centers preserves the inversion symmetry and therefore must couple only to $E_g$ vibrational modes. Furthermore, the effect is dynamic \cite{Hepp2014,Rogers2014c}, meaning that the point group of the entire system of coupled electrons and ions remains $D_{3d}$. The electron density in the ground (excited) state is given by the average of electron densities of the two $^2E_g$ ($^2E_u$) states, which is very well approximated by the density of the $e_{ux}^{2}e_{uy}^{2}e_{gx}^{3/2}e_{gy}^{3/2}$ ($e_{ux}^{3/2}e_{uy}^{3/2}e_{gx}^{2}e_{gy}^{2}$) configuration. It follows that because of its dynamic nature, the presence of the Jahn-Teller effect should have minimal effect on the results of the present Section, in particular regarding $A_{2u}$ and $E_u$ modes.

\section{Discussion \label{disc}}

In this Section we discuss how our calculations compare to experimental measurements. In a recent PL study \cite{Dietrich_NJP_2014}, which was already discussed above, two prominent phonon side peaks were observed: a narrow feature (width $w$$\approx$$5$~meV) at 63.8~meV from the ZPL and a broad one (width $w$$>$$25$~meV) at $\sim$42~meV from the ZPL. As also discussed, a clear isotope shift with respect to the ZPL line of the 63.8~meV peak was observed. At variance, the 42~meV peak did not show any noticeable change upon isotope substitution. As the authors of the paper admitted \cite{Dietrich_NJP_2014}, the lack of change could have been related to difficulties in extracting small differences in energies of broad peaks. It was also suggested \cite{Rogers2014c,Dietrich_NJP_2014} that the 63.8~meV feature was due to a $A_{2u}$ vibration of the Si atom along the defect symmetry axis.

In addition to shifts of vibronic features, significant isotopic shift of the ZPL energy was found in experiment \cite{Dietrich_NJP_2014}: $\Delta E_{28,29} = E_{28} - E_{29}=0.36$~meV, and $\Delta E_{28,30} =0.68$~meV. Typically, such shifts are explained by a difference in vibrational frequencies in the ground and the excited states \cite{Dietrich_NJP_2014,Ekimov2017}. Zero-point vibrations contribute to the $T$=$0$ value of the ZPL. As the vibrational frequencies change upon isotope substitutions, so does the ZPL. Experiments found that ZPL shifts to lower energies as the mass of the Si atom increases, and this implies that the vibrational frequencies are {\it on average} higher in the excited state than in the ground state \cite{Dietrich_NJP_2014}. Similarly large isotopic shifts were recently found for the GeV centers in diamond that are very similar to SiV centers \cite{Ekimov2017}.

Let us discuss all these experimental findings. Low-temperature luminescence lineshape reveals the vibronic structure in the ground state, so we will first discuss the vibrational modes in the ground state. On par with experiment, our calculations also indicate the existence of two resonances, one at 43.4~meV and another one at 60.1~meV. One would be tempted to claim a good agreement with experimental values of $\sim$42 and 63.8~meV. However, there is a very large discrepancy concerning the width of these resonances. While calculations indicate that both the lower-energy $A_{2u}$ resonance and the higher-energy $E_u$ resonances are narrow (widths $w$ of 1.9 and 6.1 meV, respectively), experiment indicates that the lower-energy resonance is very broad ($w>25$~meV), and only the higher-energy one is narrow ($w\approx 5$ meV). Also, in calculations frequencies of both modes undergo nearly ideal isotope shifts $\sim 1/\sqrt{m_\text{Si}}$, while in experiment only the narrow higher-energy peak shows this shift. 

The following questions arise: 1. Why both peaks are narrow in calculations, but only one in experiment? 2. Is the experimentally observed narrow 63.8~meV peak an $A_{2u}$ or an $E_u$ vibration? Density functional theory calculations work very well regarding the vibrational structure of other defects in diamond, including the NV center \cite{Gali2011b,Alkauskas_NJP_2014}, and we should expect them to work well for the SiV center as well. We therefore tentatively suggest that the experimentally observed 63.8~meV feature is in fact the $E_u$ vibration, and not the $A_{2u}$ vibration, as proposed in Ref.~\onlinecite{Dietrich_NJP_2014}. If this is indeed the case, then we conclude that experiment agrees well with calculations regarding (i) the frequency (63.8~meV vs.~60.1~meV), (ii) the width ($\sim$$5$~meV vs.\ 6.1~meV), and (iii) the isotopic shift of the resonance. We also hypothesize that the $A_{2u}$ resonance is narrow in experiment, but it does not appear in the PL spectrum; the experimental broad feature at 42~meV is in fact not related to quasi-local modes. 

However, the appearance of either $E_u$ or $A_{2u}$ mode in the experimental PL spectrum is currently not clear. Since optical transition is between $^2E_u$ and $^2E_g$ states, according to the group-theoretical analysis, in the Franck-Condon approximation one should expect to see only $A_{1g}$ phonons. As discussed above, density functional calculations indicate the presence of the dynamical Jahn-Teller effect both in the excited and the ground state \cite{Gali2013}. These two states couple to $E_g$ phonons, resulting in the $E_u \otimes E_g$ JT and $E_g \otimes E_g$ JT systems. Optical transition between these two JT systems should involve only even-symmetry ($g$) phonons, leaving the appearance of the $E_u$ mode in the experiment unexplained. This calls for alternative coupling mechanisms.

Interestingly, frequencies of $A_{2u}$ and $E_{u}$ phonons in the excited state are rather close to each other [Table \ref{Table2}]. Since these resonances have finite widths, there is some overlap in energies. Small perturbations, such as, e.g., quadratic interactions \cite{Osadko_1979} or strain, could mix these vibrations. In this case one could expect signatures of both $A_{2u}$ and $E_u$ vibrations in the emission spectrum. However, it remains unclear how such interaction could lead to the situation where only $E_u$ vibration appears, while $A_{2u}$ vibration does not (or vice versa). One more option is provided by the Herzberg-Teller effect, whereby the transition dipole moment is modulated by the vibration \cite{Osadko_1979}. When the vibration is of $A_{2u}$ symmetry, this reduces the instantaneous symmetry of our system from $D_{3d}$ to $C_{3v}$. However, degeneracy of electronic states is not removed. At variance, $E_{u}$ vibration reduces the instantaneous symmetry to $C_{1}$, removing the degeneracy and possibly leading to a appreciable modulation of the transition dipole moment. While the appearance of asymmetric modes in optically allowed transitions is very unusual \cite{Osadko_1979}, one cannot reject this possibility. The study of optical transitions and the luminescence lineshape of SiV centers is beyond the aim of the present article and requires further work. Understanding the mechanism of the vibronic coupling at SiV centers will be also pivotal in explaining experimentally measured polarization of the phonon sideband \cite{Rogers2014c}.

Now, let us discuss the experimentally observed isotope shift of the ZPL \cite{Dietrich_NJP_2014}. Our results show that the vibrational frequencies of quasi-local modes are indeed higher in the excited state than in the ground state. As discussed above, this leads to a decrease of the ZPL energy upon substitution with heavier Si atoms, in qualitative agreement with experiments. If we first assume that all of the shift is due to quasi-local modes, we could estimate the change of the ZPL when $^{28}$Si is replaced by $^{29}$Si as
\begin{eqnarray}
\Delta E_{28,29} \approx \frac{1}{2}\hbar\left(\Omega^e_{A_{2u}} + 2\Omega^e_{E_u} - \Omega^g_{A_{2u}} - 2\Omega^g_{E_u}\right) \nonumber
\\ 
\times \left(1-\sqrt{\frac{m_{28}}{m_{29}}}\right).
\end{eqnarray}
The estimate yields a value of 0.17~meV, almost twice smaller than the experimental value of 0.36~meV. Similarly, the calculated value for $\Delta E_{28,30}$ of 0.32~meV is smaller than the experimental one of 0.68~meV. One could alternatively calculate the shift by including the contribution of all vibrations. For this purpose, we can determine the contribution of zero-point vibrations to the energies of excited and ground states as:
\begin{equation}
E_\text{ZPV} = \frac{1}{2}\sum_i \hbar\omega_i^e - \frac{1}{2}\sum_i \hbar\omega_i^g. 
\label{ZPV}
\end{equation}
The sum converges very fast as a function of the supercell size, and we obtain values 17.06~meV for $^{28}$Si, 16.80~meV for $^{29}$Si, and 16.56~meV for $^{30}$Si. This gives $\Delta E_{28,29}$=0.26~meV and $\Delta E_{28,30}$=0.50~meV. These results are much closer to the experimental values. It is likely that in order to explain the remaining discrepancy one would need to consider the above-mentioned Jahn-Teller effect, in which the zero-point vibrational contributions differ from the one given in Eq.~(\ref{ZPV}).

\section{Summary and Conclusions \label{summ}}
In summary, we have studied vibrational spectrum of the negatively charged silicon-vacancy center in diamond by first-principles density functional theory calculations. The electronic structure, geometry, and force constants were calculated with the hybrid density functional of Heyd, Scuseria, and Ernzerhof. The vibrational spectrum of the defect was modeled in large supercells containing up to 5832 atoms (nominal size) using the embedding procedure whereby the dynamical matrix of a large supercell was constructed from the dynamical matrix of a defect in a 216-atom supercell and the dynamical matrix of bulk diamond. We find that even-symmetry vibrations ($A_{1g}$, $A_{2g}$, and $E_g$), as well $A_{1u}$ vibrations are essentially perturbed bulk modes. However, we find very pronounced quasi-local modes of $A_{2u}$ and $E_u$ symmetries that correspond to the vibration of the Si atom along and perpendicular to the defect symmetry axis, respectively. We have presented the methodology to calculate isotope shifts of these modes, and find excellent agreement with experiment. We have also found that the vibrational frequencies in the electronic excited state are larger than in the ground state, also in full agreement with experimental findings. Finally, we suggest that the experimental observed feature in the photoluminescence spectrum at 63.8~meV is a quasi-local $E_u$ mode. While the appearance of this feature in the experimental spectrum remains unexplained and calls for further investigations, we provided some possible physical mechanisms. By systematically addressing the vibrational properties of the SiV center, our work will be helpful in understanding physical properties of single-photon emitters in diamond, in particular the GeV center \cite{Ekimov2017} and the SnV center \cite{Iwasaki2017}, as well as vibrational resonances in general. 

\acknowledgments
We thank M. Bijeikyt\.{e}, M. W. Doherty, F. Jelezko, A. Matulis, and I. S. Osad'ko for fruitful discussions. A.~G. acknowledges the support from NIIF Supercomputer Center Grant No.~1090, and National Research Development and Innovation Office of Hungary (NKFIH) within the Quantum Technology National Excellence Program (Project No.~2017-1.2.1-NKP-2017-00001).
G.~T. acknowledges support \'{U}NKP-17-3-III New National Excellence Program of the Ministry of Human Capacities of Hungary.
L.~R. and A.~A. were supported by a grant No.~M-ERA.NET-1/2015 from the Research Council of Lithuania. Additional calculations were performed at the High Performance Computing Center ``HPC Saul\.etekis'' in Faculty of Physics, Vilnius University.


\begin{thebibliography}{32}
\bibitem [{IRR()}]{IRREP}%
  \BibitemOpen
  \href@noop {} {}\bibinfo {note} {Capital letters are used to label
  irreducible representations of many-electron wavefunctions (e.g., $^2E_g$)
  and vibrations (e.g., $A_{2u}$), while small letters (e.g., $e_u$ or
  $a_{1g}$) are used for single-electron states.}\BibitemShut {Stop}%
\end{thebibliography}
\end{document}